\documentclass[a4paper]{jpconf}
\usepackage{graphicx}

\usepackage{lineno}

\begin{document}
\title{Bridging Worlds: Achieving Language Interoperability between Julia and Python in Scientific Computing}

\author{Ianna Osborne$^1$, Jim Pivarski$^1$, Jerry Ling$^2$}

\address{$^1$ Princeton University, Princeton, NJ 08544, USA}
\address{$^2$ Harvard University, Massachusetts Hall, Cambridge, MA 02138, USA}

\ead {ianna.osborne@cern.ch}

\begin{abstract}
In the realm of scientific computing, both Julia and Python have established themselves as powerful tools. Within the context of High Energy Physics (HEP) data analysis, Python has been traditionally favored, yet there exists a compelling case for migrating legacy software to Julia. This article focuses on language interoperability, specifically exploring how Awkward Array data structures can bridge the gap between Julia and Python. The talk offers insights into key considerations such as memory management, data buffer copies, and dependency handling. It delves into the performance enhancements achieved by invoking Julia from Python and vice versa, particularly for intensive array-oriented calculations involving large-scale, though not excessively dimensional, arrays of HEP data.
The advantages and challenges inherent in achieving interoperability between Julia and Python in the domain of scientific computing are discussed.

\end{abstract}

\section{Introduction}

Python has long been favored among scientists and researchers for its simplicity and extensive ecosystem. In scientific computing, both Python and Julia are widely utilized for their ease of use, powerful libraries, and growing capabilities. While Python remains popular, Julia is gaining traction, especially for its speed and flexibility.

The interoperability between Python and Julia is being discussed in the presentations such as those cited~\cite{jim-ref},~\cite{julia-ref}. They highlight initiatives aimed at bridging the gap between these two languages. Additionally, tools like PythonCall and JuliaCall~\cite{pythoncall-ref} provide symmetric interfaces for calling code between Python and Julia.

However, working with data across language boundaries can pose challenges, with data format conversion often being the most difficult aspect. Awkward Array~\cite{awkward-ref} used by physicists for data manipulation and analysis is one of the tools in the scientific Python ecosystem that addresses this issue by its design.

\section{Interoperability}

PythonCall and JuliaCall provide an easy way to invoke Python code from Julia and Julia code from Python. This bidirectional communication channel facilitates data sharing between the two languages, enabling users to leverage the strengths of both Python and Julia in their scientific computing workflows.

There have been various attempts to establish connectivity between Julia and Python languages. Currently, there are 14 repositories or projects dedicated to addressing this specific need. The decision to use PythonCall and JuliaCall was based on their active maintenance and ongoing development, ensuring reliable and up-to-date functionality.

One of the key challenges in this two language interoperability is the Global Interpreter Lock (GIL) in Python, which can impact performance when calling Julia from Python. Efforts are underway to mitigate this issue, with future versions of JuliaCall expected to offer solutions for releasing the GIL when calling Julia code from Python.

Notably, when calling Julia from Python using JuliaCall, it's possible to drop the GIL. As the documentation mentions: 'If you can ensure your Julia code doesn't call back into Python, you can release the GIL yourself.' A forthcoming version of JuliaCall aims to simplify this process with functions like $`some\_julia\_function.jl\_call\_nogil(x, y, z)`$.

\section{Data exchange between Python and Julia}

While PythonCall and JuliaCall facilitate Awkward Array data exchange between Python and Julia, an additional layer is needed to pass data structures between the languages. The Awkward Array design simplifies this process through its 'to\_buffers' and 'from\_buffers' functions, which enabling access to the same memory of Awkward Array objects. Ultimately, it is AwkwardArray.jl package that enables this memory access, thereby allowing smooth data exchange between Python and Julia.

\begin{figure}[htp]
    \centering
    \includegraphics[width=16cm]{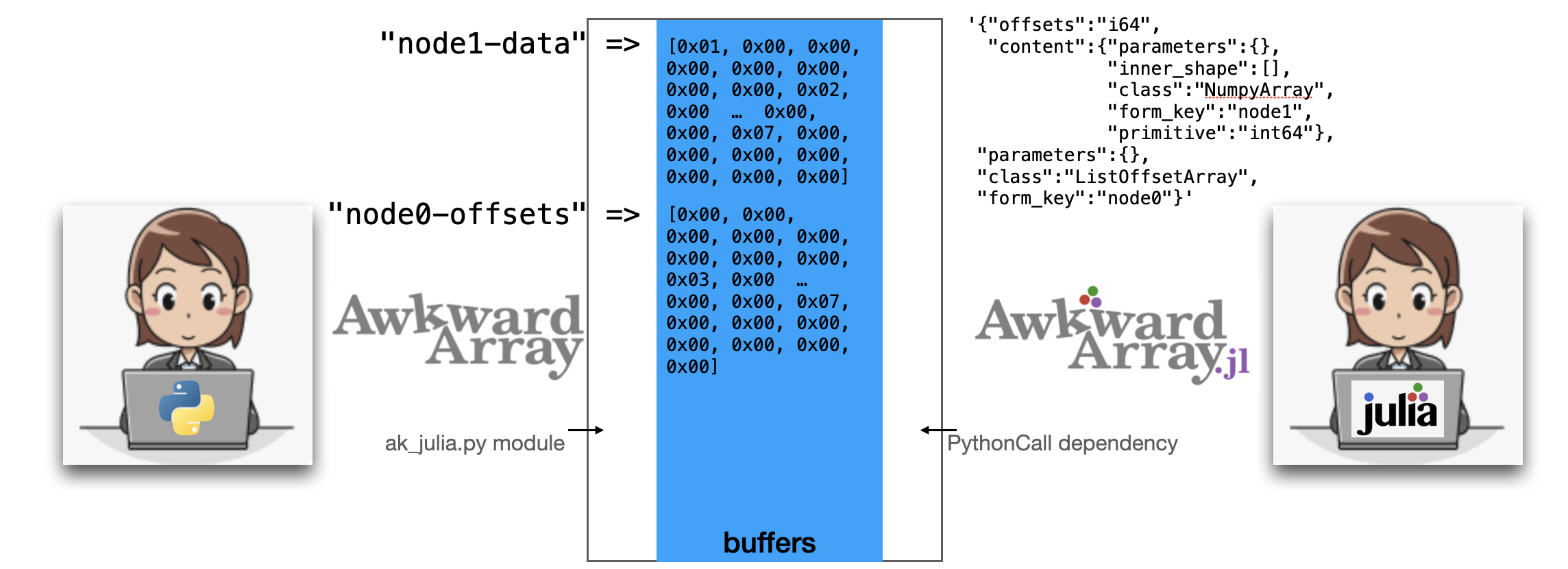}
    \caption{Awkward Array data sharing across language boundaries.}
    \label{fig:AwkwardArray_data}
\end{figure}

\section{Converting Awkward Arrays from Julia to Python and from Python to Julia}

The process of passing data structures from Python to Julia and utilizing Awkward Array establishes a bridge between the two languages, enabling bidirectional interoperability.

AwkwardArray implements a convert function that, given an AwkwardArray type (Primitive Array in Julia, which is equivalent to Numpy Array in Python), converts it to an Awkward Array in Python. Similarly, PythonCall provides a pycovert function that converts from a known Python type to a specified type.

\begin{verbatim}
using PythonCall
using AwkwardArray: convert
\end{verbatim}

Here is an example how to create a Julia awkward array, convert it to a Python awkward array, and check if the function returns a Python awkward array, and also check if the awkward array has the correct layout.

\begin{verbatim}
array = AwkwardArray.ListOffsetArray(
    [0, 3, 3, 5],
    AwkwardArray.PrimitiveArray([1.1, 2.2, 3.3, 4.4, 5.5]),
)
py_array = convert(array)
py_array isa Py
typeof(py_array) == Py
\end{verbatim}

We can also use:

\begin{verbatim}
ak_array = pyconvert(Vector, pyimport("awkward").to_list(py_array))
ak_array == [[1.1, 2.2, 3.3], [], [4.4, 5.5]]
\end{verbatim}

Let's create a Python awkward array, convert it to Julia and check:
\begin{verbatim}
py_array = pyimport("awkward").Array([[1.1, 2.2, 3.3], [], [4.4, 5.5]])

array = convert(py_array)
array isa AwkwardArray.ListOffsetArray

array == [[1.1, 2.2, 3.3], [], [4.4, 5.5]]
\end{verbatim}
   
\section{Runtime Environment and Julia Package Manager}

Ensuring compatibility between Python and Julia environments is crucial for integration. The Conda.jl package enable users to utilize Conda as a cross-platform binary provider for Julia and other Julia packages. This is particularly beneficial when dealing with complex dependencies, such as those required by Python libraries.

When working with Julia, its package manager maintains a consistent and reliable runtime environment. Julia's project required packages are described in the Project.toml file, that specifies and installs dependencies. Using both Python and Julia together necessitates operating within the same environment to ensure that package versions are aligned.

By leveraging Conda.jl and Julia's package manager, users can  manage dependencies and create runtime environments that support the interoperability of Python and Julia.

\section{Python Module JuliaCall}

While Conda offers a comprehensive repository of packages, not all projects are available through this platform. For instance, JuliaCall must be installed using pip due to its specific requirements.

The Python module JuliaCall provides a straightforward way to establish communication between Python and Julia environments. Installing JuliaCall is as simple as running `pip install juliacall`.

Once installed, JuliaCall interacts with the Julia package manager directly from the Python REPL. It takes care of installing essential packages like PythonCall, ensuring a smooth setup process.

\section{Runtime Environment and Python Packages}

On the Python side, the runtime environment plays a crucial role in ensuring smooth interoperability with Julia. While I won't dive into the specifics here, it's essential to address version compatibility issues. Coordinating different versions of Python and Julia, along with their respective packages, can present challenges that need careful navigation.

\section{Dependency management}

Dependency management is another important consideration. When using both Python and Julia, it's essential to ensure that all dependencies play nice together. That means making sure everything works smoothly in both Python and Julia environments.

For instance, Python packages like Awkward Array can be installed using Conda for Python, but ensuring compatibility with Julia adds an extra layer of complexity. A robust environment is created by carefully managing dependencies and versioning, so that interoperability between Python and Julia is assured, enabling resource sharing across both ecosystems.

\section{Performance Benefits}

Calling Julia from Python can offer significant performance benefits for running analysis code. However, it's crucial to follow best practices to maximize efficiency and maintainability. When integrating Julia code into your Python workflow, it's advisable to encapsulate performance-critical operations within functions. These functions should accept arguments rather than relying on global variables -- this is essential for Julia compiler to emit performant code, and it also promots modularity and reusability.

Let's consider an example demonstrating how to leverage Julia for calculating path lengths in an Awkward Array within a Python environment.

\begin{verbatim}
function path_length(array)
    total = 0.0
    for i in eachindex(array)
        for j in eachindex(array[i])
            total += array[i][j]
        end
    end
    return total
end
\end{verbatim}

It's essential to keep in mind Julia's performance recommendations. To ensure optimal performance, it's advisable to avoid using containers with abstract type parameters and minimize reliance on object-oriented programming (OOP) paradigms for performance-sensitive code. Following these guidelines, one can leverage PythonCall effectively while maintaining high-performance standards in their Julia code.

\section{Conversion Rules}
When converting Python objects to Julia objects using PythonCall, custom conversion rules are defined. These rules dictate how Python types are translated into their corresponding Julia equivalents. Specifically, when dealing with complex data structures like Awkward Array, PythonCall allows us to define conversion rules tailored to our specific needs.

\begin{verbatim}
function pyconvert_rule_awkward_array_listoffset(
  ::Type{AwkwardArray.ListOffsetArray}, x::Py)
    array = AwkwardArray.convert(x)
    return PythonCall.pyconvert_return(array)
end
\end{verbatim}

Once the custom conversion rules for translating Python objects to Julia equivalents are defined, the next step is to register them with PythonCall. This registration process ensures that PythonCall recognizes the custom rules and applies them appropriately during conversions.

Registering a custom conversion rule with PythonCall typically involves specifying a conversion function along with relevant metadata to identify the types involved. Registering the rule and integrating it into PythonCall's conversion pipeline a smooth and an efficient translation of Python objects to Julia types is assured.

\begin{verbatim}
PythonCall.pyconvert_add_rule(
    "awkward.highlevel:Array", 
    AwkwardArray.ListOffsetArray, 
    pyconvert_rule_awkward_array_listoffset, 
    PythonCall.PYCONVERT_PRIORITY_ARRAY
)
\end{verbatim}
    
This integration process not only enhances interoperability between Python and Julia but also ensures that our custom conversion logic is consistently applied whenever Python objects are passed to Julia via PythonCall. Thus, by registering our conversion rule with PythonCall, we can effectively bridge the gap between the two languages while maintaining flexibility and performance.

\section{Potential Issues}

When working with Awkward Arrays across Julia and Python, it's essential to be aware of potential issues that may arise, particularly regarding indexing and field selection. By being mindful of these potential issues and addressing them proactively, we mitigate compatibility challenges and ensure smooth integration between Awkward Arrays in Julia and Python.

Consider an example where we create an Awkward Array in Julia and then convert it to an Awkward Array in Python. While the syntax for both languages is generally straightforward, subtle differences in indexing and field selection could lead to unexpected behavior or errors.

\begin{verbatim}
>>> jl.content_layout[0]
Julia: {a: 1, b: 1.1}
>>> jl.content_layout[2]
Julia: {a: 3, b: 3.3}
>>> jl.content_layout[3]
Julia: {a: 4, b: 4.4}
>>> jl.seval("""
... content_layout[0]
... """)
juliacall.JuliaError: BoundsError: attempt to access 
5-element Vector{Int64} at index [0]
\end{verbatim}

For instance, certain indexing conventions or field selection methods may differ between the two languages, requiring careful attention to ensure consistent results. Additionally, variations in how data is represented or accessed internally could impact the interoperability of Awkward Arrays between Julia and Python.

Addressing potential issues related to indexing conventions, it's important to recognize that Julia adopts a 1-based indexing system, unlike some other programming languages that utilize a 0-based indexing convention.

To ensure code portability and compatibility across different environments, it's advisable to refrain from hardcoding numerical indices when accessing elements within containers. Instead, developers can leverage the `firstindex` function, which dynamically adjusts the indexing behavior based on the language's conventions.

By utilizing `firstindex` to determine the starting index for container access, code remains agnostic to the specific indexing convention of the underlying language. This approach enhances code flexibility and facilitates transition between different programming environments, minimizing potential errors or inconsistencies arising from indexing discrepancies.

When dealing with Record Arrays, another potential issue arises concerning the handling of fields. One approach to mitigate this challenge is by constraining the interface of the Record Array, perhaps through the implementation of a 'field' attribute.

By providing a designated 'field' attribute, users can explicitly specify which fields they intend to access or manipulate within the Record Array. This approach reduces ambiguity regarding field selection, thereby enhancing the clarity and predictability of the code.

Implementing a 'field' attribute also promotes consistency across different operations involving Record Arrays, simplifying the development process and reducing the likelihood of errors or misunderstandings.

\begin{verbatim}
>>> jl.seval("""
content_layout[:a]
""")
Julia:
5-element AwkwardArray.PrimitiveArray{Int64, Vector{Int64}, :default}:
 1
 2
 3
 4
 5
>>> py_rec_arr = jl.convert(jl.content_layout)
>>> py_rec_arr["a"]
<Array [1, 2, 3, 4, 5] type='5 * int64'>
\end{verbatim}

Overall, by incorporating such design considerations and providing clear guidelines for field access and manipulation, developers can improve the usability and robustness of Record Arrays in both Julia and Python environments.

Embracing a more adaptive and language-agnostic indexing strategy not only promotes code portability but also fosters collaboration and interoperability across diverse scientific computing ecosystems. It empowers developers to write more versatile and resilient code that can readily adapt to varying language conventions and environments, ultimately enhancing the efficiency and effectiveness of scientific computing workflows.

\section{Use Cases for Integrating Julia and Python}

Exploring the use cases for integrating Julia and Python highlights the versatility and synergistic potential of these languages in scientific computing and beyond:

\begin{itemize}
    \item Integrating high-performance Julia libraries into existing Python projects

    \item Leveraging Python's extensive data analysis and visualization libraries in Julia workflows for data manipulation, exploration, and visualization

    \item Building hybrid applications for scientific simulations, machine learning models, or numerical algorithms

    \item Adding fast Julia libraries to Python projects for addressing performance bottlenecks and handling computationally intensive tasks

    \item Creating apps that combine ease of use with computational speed to deliver a productive user experience without sacrificing performance
\end{itemize}

In essence, the integration of Julia and Python opens up the possibilities for tackling complex scientific computing tasks.

\section{CMS Data Analysis Example}

Let's delve into a practical example of how the integration of Julia and Python can be applied in a real-world scenario, focusing on data analysis within the context of the Compact Muon Solenoid (CMS) experiment at CERN.

\begin{verbatim}
function invariant_mass(tree)
    layout = AwkwardArray.PrimitiveArray{Float64}()
    for event in tree
        event.nMuon == 2 || continue
        if event.Muon_charge[1] == event.Muon_charge[2]
            continue
        end
        result = sqrt(2 * event.Muon_pt[1] * event.Muon_pt[2] *
        (cosh(event.Muon_eta[1] - event.Muon_eta[2]) -
        cos(event.Muon_phi[1] - event.Muon_phi[2])))
        if result > 70
            push!(layout, result)
        end
    end
    layout
end
\end{verbatim}

The CMS experiment generates vast amounts of particle collision data~\cite{cms-data}, which require sophisticated analysis techniques to extract meaningful insights. Here's how the combined power of Julia and Python can streamline this process:

\begin{itemize}
    \item Data Preprocessing with Python: Python's extensive ecosystem of data processing libraries, such as Awkward Array, uproot~\cite{uproot-ref}, Pandas and NumPy, excels at handling large datasets efficiently. Researchers can use Python to preprocess raw data from the CMS experiment, performing tasks such as data cleaning, transformation, and feature engineering.
    \item High-Performance Analysis with Julia: Once the data has been preprocessed, researchers can leverage Julia's high-performance computing capabilities to perform complex analyses and simulations. Julia's speed and efficiency make it well-suited for tasks such as fitting models, running simulations, and performing statistical analyses on large datasets.
    \item Visualization and Reporting with Python: After conducting the analysis in Julia, researchers can use Python's robust visualization libraries to create insightful visualizations and reports. Python's flexibility and ease of use make it ideal for generating publication-quality plots and sharing results with colleagues.
    \item Iterative Development and Collaboration: Throughout the analysis process, researchers can transition between Python and Julia, leveraging the strengths of each language as needed. This iterative development approach fosters collaboration and allows researchers to experiment with different analysis techniques and algorithms rapidly.
\end{itemize}

By combining the strengths of Julia and Python, researchers can streamline the data analysis workflow for the CMS experiment, enabling faster insights and discoveries in particle physics research.

\section{Conclusion}

The integration of Julia and Python offers a powerful solution for tackling complex scientific computing tasks, enabling researchers and developers to harness the strengths of each language to drive innovation and discovery. By combining Python's ease of use and extensive ecosystem with Julia's high-performance computing capabilities, physicists can create versatile, efficient, and scalable solutions for a wide range of applications. Whether developing sophisticated simulations, analyzing large datasets, or building interactive visualizations, the combined power of Julia and Python helps users to push the boundaries of scientific computing and accelerate the pace of discovery.

\section{Acknowledgment}
This work is supported by NSF cooperative agreement OAC-1836650 (IRIS-HEP) and NSF cooperative agreement PHY-2121686 (US-CMS LHC Ops).


\section*{References}
\begin{thebibliography}{9}

\bibitem{awkward-ref}
Pivarski, J., Osborne, I., Ifrim, I., Schreiner, H., Hollands, A., Biswas, A., Das, P., Roy Choudhury, S., Smith, N., Goyal, M. (2018). Awkward Array [Computer software]. \verb"https://doi.org/10.5281/zenodo.4341376"

\bibitem{jim-ref} Pivarski, J. Engaging the HEP community in Julia, JuliaHEP2023 Workshop \verb"https://indi.to/cnxwK"

\bibitem{julia-ref} Gál, T., Is Julia ready to be adopted by HEP?, JuliaHEP2023 Workshop \verb"https://indi.to/sJCcj"

\bibitem{awkward-julia-ref} AwkwardArray.jl [Computer software]. \verb"https://github.com/JuliaHEP/AwkwardArray.jl"

\bibitem{pythoncall-ref} Rowley, Christopher, (2022) PythonCall.jl: Python and Julia in harmony [Computer software]. \verb"https://github.com/JuliaPy/PythonCall.jl"

\bibitem{uproot-ref} Pivarski, J., Schreiner, H., Hollands, A., Das, P., Kothari, K., Roy, A., Ling, J., Smith, N., Burr, C., Stark, G. (2017). Uproot [Computer software]. \verb"https://doi.org/10.5281/zenodo.4340632"

\bibitem{unroot-ref} Gál, T., Ling, J., Amin, N. (2021). UnROOT: an I/O library for the CERN ROOT file format written in Julia (Version v1) [Computer software]. \verb"https://doi.org/10.21105/joss.04452"

\bibitem{awkward-rdf-pyhep2022-ref} Osborne, I., Pivarski, J., (2022, September 15). Awkward RDataFrame Tutorial. PyHEP 2022 (virtual) Workshop. Zenodo. \verb"https://doi.org/10.5281/zenodo.7081586"

\bibitem{cms-data} CERN Open Data portal \verb"DOI:10.7483/OPENDATA.CMS.LVG5.QT81" (CMS data)

\end{thebibliography}
\end{document}